\newcommand{\ds}{\displaystyle}
\documentstyle[12pt,epsf]{article}
\begin{document}

\begin{center}
{\bf Transient Dynamics in Magnetic Force Microscopy  for a Single-Spin
 Measurement}\\ \ \\

 {G.P. Berman,$\!^1$ F. Borgonovi,$\!^{2}$ G. V.
L\'opez,$\!^3$ V.I. Tsifrinovich~$\!^{4}$}\\ \ \\
\end{center}  

\begin{center}
{$^1$Theoretical Division and CNLS,
Los Alamos National Laboratory, Los Alamos, NM 87545}\\
 
{$^2$Dipartimento di Matematica e Fisica, Universit\`a Cattolica,
via Musei 41 , 25121 Brescia, Italy, and I.N.F.M.,  Unit\`a
di Brescia  and I.N.F.N., sezione di Pavia }\\

{$^3$
Departamento de F\'{\i}sica de la Universidad de Guadalajara
 S.R. 500, 44420
Guadalajara, Jalisco, M\'exico}\\

{$^4$
IDS Department, Polytechnic University, Six Metrotech Center, Brooklyn, 
New York 11201}
\end{center}

%\maketitle
\begin{abstract}
\noindent
We analyze a single-spin measurement using a transient process in magnetic 
force microscopy (MFM) which could increase
the maximum operating temperature by a factor of $Q$ (the quality factor of 
the 
cantilever) in comparison with the static Stern-Gerlach effect. We obtain an 
exact solution of the master 
equation, which confirms this result. We also discuss the conditions
required to create a macroscopic Schr\"odinger cat state in the cantilever.
\end{abstract}

\section{\bf  Introduction}
The routine magnetic force microscopy (MFM) seems to be of a little 
use for a single spin
detection in solids. Even for unrealistically small distances between 
the ferromagnetic
particle on the cantilever tip and the spin, the maximum temperature for a 
single-spin
measurement comes to a millikelvin region [1]. In this paper, we show that 
using a
transient process, one can increase the maximum temperature of a single-spin 
measurement
by a factor of $Q$ (the quality factor of the cantilever). Alternatively, one 
can increase the distance between the ferromagnetic
particle and the cantilever tip if one is willing to work at millikelvin 
temperatures.\\ \ \\
\noindent
In section 2 we explain the basic idea of our work. In sections 3-5 we obtain 
an exact solution
of the master equations, which confirms our idea. In section 6 we derive the 
conditions
for creating a macroscopic Schr\"odinger cat state (MSCS) in MFM. 
In Conclusion
we summarize our results.\\ \ \\
\section{\bf Transient Process in MFM}
In this section we describe the basic idea of our work. Suppose that a 
ferromagnetic particle on the 
cantilever tip interacts with a single spin on the solid surface. 
(See Fig. 1.) The equilibrium position
of the cantilever tip depends on the spin direction. The distance 
between two possible
equilibrium positions, corresponding to two spin stationary states, 
is given by $2F/k_c$, 
where $k_c$ is the cantilever spring constant, $F$ is the magneto-static 
force between the 
ferromagnetic particle and the spin. In order to measure the state of the 
spin, this distance
must be greater than twice the uncertainty due to the 
thermodynamical noise of the 
cantilever position. This uncertainty can be estimated 
as $(k_{_B}T/k_c)^{1/2}$, where $k_{B}$
is the Boltzmann's constant and $T$ is the temperature.
Thus, the condition for a single spin measurement 
(the static Stern-Gerlach effect) is
\cite{g}
\begin{equation}
T<T_{max}=F^2/k_{_B}k_c.
\end{equation}
Now, assume that we quickly change the stationary state of the spin and 
consider the transient
cantilever vibrations after this change. The amplitude of the cantilever 
vibrations at 
time $t\ll Q/\omega_c$ ($\omega_c$ is the cantilever frequency 
and $Q/\omega_c$ is
the time constant of the cantilever) is $4F/k_c$. In addition, assume 
that we detect the 
position and momentum of a point on the cantilever tip with an accuracy 
that satisfies the quantum limit
$(\delta P_{_Z})(\delta Z)\approx\hbar/2$ (the cantilever oscillates 
along the z-axis.) To 
find out the state of the spin
we are going to compare the observed trajectory of the cantilever 
tip with the theoretical
prediction. The theory predicts the cantilever trajectory
within an uncertainty due to the thermal noise. However, the thermal 
uncertainty of the cantilever
position at time $t\ll Q/\omega_c$ increases as  
$[t(\omega_c/Q)(k_{_B}T/k_c)]^{1/2}$.
We can obtain this expression assuming an initial ($t\ll Q/\omega_c$) 
thermal diffusion 
with two common properties: 1) The dispersion $\overline{(\delta Z)^2}$ 
is proportional to time, $t$, and 2) The uncertainty of the cantilever 
position equals its thermodynamical 
value if we formally put $t=Q/\omega_c$. \\ \\
At the time $t=\pi/\omega_c$ (half of the first
period), the distance between two possible cantilever positions takes 
its maximum possible 
value $4F/k_c$. At the same time, the thermal uncertainty of the position 
predicted by
the theory is still much smaller than its thermodynamical value. Now, the 
condition for
a single spin measurement is
\begin{equation}
T<T_{max}={ 4QF^2\over\pi k_{_B}k_c}.
\end{equation}
One can see that the maximum temperature for a single spin measurement 
increases by a factor
of $Q$, compared with the static Stern-Gerlach effect.\\ \ \\
In the next three sections, we will confirm this estimation by direct 
solution of the master
equation.
\section{\bf The Hamiltonian and the Master Equation}
We assume that the cantilever tip with an attached ferromagnetic particle 
can oscillate along the z-axis.
(See Fig. 1.) A single paramagnetic atom with spin $1/2$ is placed near the 
cantilever tip. 
The dimensionless Hamiltonian of the cantilever tip interacting with a 
single spin is

\begin{equation} 
\hat H={1\over 2}(\hat p_z^2+ \hat z^2)-2\eta \hat z \hat S_z.
\label{eqh}
\end{equation}
We introduced the following notation (below we omit hats for operators): 

\begin{equation}
z=Z/Z_q,\quad p_z=P_z/P_q,\quad 
\eta=g\mu_{_B}\left|{\partial B_z\over\partial z}\right|/2F_q\ ,
\end{equation}
where $Z_q$, $P_q$ and $F_q$ are the ``quants" of the 
coordinate, momentum and force
acting on the cantilever,

\begin{equation}
Z_q=\left({\hbar\omega_c\over k_c}\right)^{1/2},\quad P_q=\hbar/Z_q,\quad 
F_q=k_cZ_q .
\end{equation}
The variables $Z$ and $P_z$ are the ``dimensional" coordinate and 
momentum of the 
cantilever tip, $k_c$ is the 
cantilever ``spring constant", $\omega_c$ is its frequency, 
$g$ is the ``g-factor" of the 
spin (below we use $g=2$), $\partial B_z/\partial z$ is the 
magnetic field gradient  
produced by the ferromagnetic particle at the spin location
when the cantilever is in the equilibrium position with  no spin 
($z=0$). Note that the 
cantilever interacts with the z-component of the spin, which is an 
integral of motion
in our system. In the Hamiltonian (3) we omitted the term 
($g\mu_{_B}B_0/\hbar\omega_c)\hat S_z$,
where $B_0$ is the magnetic field on the spin when the cantilever 
is at the origin ($z=0$).
This term may be eliminated ``physically'' (by application of a 
uniform external field of
magnitude $B_0$ in the negative $z$-direction) or ``mathematically'' 
(by transferring to the 
system of coordinates rotating with the frequency $g\mu_{_B}B_0/\hbar$).  
\\ \ \\
The master equation describes the evolution of the density matrix of the 
system 
interacting with the environment (see for example [2-5]). 
We are taking into account the interaction of the cantilever with its 
environment, and ignore the
direct interaction between the spin and the environment, assuming that 
the spin relaxation and decoherence times are large enough. The effect of the 
environment depends on its ``spectral density", i.e. the density of 
environmental
oscillators at a given frequency. Probably, the simplest model of the 
environment
is the ``ohmic" model, where the spectral density is proportional to the 
frequency
$\omega$ for $\omega<\Omega$, where $\Omega$  is the cutoff frequency 
for the environment.
A master equation for the non-ohmic environment has been derived in [5]. 
For the ohmic model, 
the simplest master equation has been obtained in [2]. This equation is 
valid in the
``high temperature limit" $k_{_B}T>>\hbar\Omega$. The master equation 
derived in [3] is valid
for arbitrary temperature. As pointed out in [5], both equations [2] 
and [3] fail at times shorter than or close to $\hbar/k_{_B}T$.\\ \ \\
We are going to consider the ``gedanken experiment" discussed in \cite{g}. 
Suppose that initially 
($t=0$) the spin is in a superposition of the two states 
with the $z$-projection of the spin $S_z=1/2$ and
$S_z=-1/2$. These two states of the spin correspond to two different 
equilibrium positions 
of the cantilever tip. Thus, the cantilever (without decoherence)  would 
transform
into a MSCS: two simultaneous equilibrium positions. Certainly, decoherence
will destroy this state. The master equation describes both the appearance
of the MSCS and its destruction due to decoherence.\\ \ \\
Following \cite{g}, we consider the ultra-thin cantilever reported on [6]. 
It has the spring
constant $k_c=6.5\times 10^{-6}~N/m$, the frequency,  $\omega_c/2\pi=1.7~kHz$,
 and the
quality factor, $Q=6700$. The ferromagnetic particle on the cantilever tip is 
taken
to be a sphere of radius $R=15~nm$ at a distance $5~nm$ from the paramagnetic 
atom. (Below we consider conditions for increasing the distance between the 
cantilever and spin.)
For this case, the static displacement of the cantilever tip due to its 
interaction
with the single spin exceeds the thermal vibrations of the cantilever for 
 temperatures

\begin{equation}
T<T_{max}={(\mu_{_B}\partial B_z/\partial z)^2\over k_{_B}k_c}\approx 1.7~mK\ .
\end{equation}
In our gedanken experiment for the temperature $T>>\hbar\omega_c/k_{_B}\approx 
8\times 10^{-8}~K$, we can use the simplest high temperature limit in 
the ``ohmic 
model". \\ \ \\
The master equation in the high temperature limit can be written in the 
form [2]

\begin{equation}
\begin{array}{lll}
 {\ds\partial\rho_{s,s'}\over\ds\partial \tau} &=\Biggl[{\ds i\over 2}
    \left(\partial_{zz}-\partial_{z'z'}\right)-{\ds i\over\ds 2}(z^2-z'^2)
    -{\ds 1\over\ds 2} \beta (z-z')(\partial_z-\partial_{z'})
    -D\beta(z-z')^2\\
  & -2 i\eta(z's'-zs)\Biggr]\rho_{s,s'}.
\label{eqr}
\end{array}
\end{equation}
Here, $s$ and $s'$ take values $\pm 1/2$ (we use $s$ instead of $S_z$), 
$\tau=\omega_ct$, 
$\beta=1/Q$ and $D=k_{_B}T/\hbar\omega_c$.
Using new coordinates
\begin{equation}
 r =z-z',\quad\quad
 R = \frac{1}{2} (z+z'). 
\label{nco}
\end{equation}
Eq. (\ref{eqr}) can be  written as:
\begin{equation}
\begin{array}{lll}  
&{\ds\partial\rho_{s,s'}(R,r,\tau)\over\ds\partial \tau}=\\ \\
&\biggl\{i\partial_{Rr}-i Rr-
\beta r\partial_r-D\beta r^2 
 -i\eta\biggl[(2R-r)s'-(2R+r)s \biggr] \biggl\}~\rho_{s,s'}(R,r,\tau).
\end{array}
\label{eqr1}
\end{equation} 
Performing a Fourier transformation of this equation with respect to 
the variable  ``R'', one obtains, after re-arrangements,
\begin{equation}   
{\partial\hat\rho_{s,s'}\over\partial \tau}+(\beta r-k)
{\partial\hat{\rho}_{s,s'}\over
\partial r}+
    \biggl[
    r+2\eta(s'-s)\biggr]{\partial\hat{\rho}_{s,s'}\over\partial k}=
    \biggl[-D\beta r^2+i\eta r(s'+s)\biggr]
\hat\rho_{s,s'},
\label{eqr2}
\end{equation}
where
 
$$
\hat\rho_{s,s'}(k,r,\tau)=\int_{-\infty}^{+\infty}e^{ikR}
\rho_{s,s'}(R,r,\tau)~dR
.
$$
We can study separately the spin diagonal case ($s=s'$) and the 
off--diagonal case
($s\ne s'$). For $s'=s$ (up-up or down-down spins), we have
the following equation:
\begin{equation}
{\partial\hat\rho_{s,s}\over\partial \tau}+(\beta
r-k){\partial\hat\rho_{s,s}\over\partial r}+
r{\partial\hat\rho_{s,s}\over\partial k}=\biggl(-D\beta r^2+ 2 i\eta r
s\biggr)~\hat\rho_{s,s},
\label{eqrd}
\end{equation}
and for  $ s' \ne s$ (up-down or down-up spins):

\begin{equation}
{\partial\hat\rho_{s,-s}\over\partial \tau}+(\beta r-k)
{\partial\hat\rho_{s,-s}\over\partial r}+
( r+4\eta s){\partial\hat\rho_{s, -s}\over\partial k}=
 -D\beta r^2 \ \hat\rho_{s,-s}.
\label{eqrnd}
\end{equation}
We will derive the exact solution of the master equation (\ref{eqr}) 
for the case when
the spin is ``prepared" initially in the superposition of two states 
with $s=1/2$ and
$s=-1/2$, while the cantilever tip is in the quasiclassical coherent state
\begin{equation}
\psi(z,s,0)= \frac{1}{(\pi)^{1/4}}
\exp [ ip_0z - (z-z_0)^2/2 ] \otimes\left(\begin{array}{c}a\\ \\  
b\end{array}\right),
\label{psi}
\end{equation}
where the amplitudes $a$ and $b$ correspond to the values of 
$s=1/2$ and $s=-1/2$ respectively. The corresponding density 
matrix can be written as

\begin{equation}
\rho_{ss'}(z,z',0)=\psi(z,s,0)\otimes\psi^{\dag}(z',s',0).
\end{equation}
Note that we consider an ensemble of spin-cantilever systems with 
the same initial
state. This implies that the experimenter can detect the position 
and momentum of
a point  on the cantilever tip with quantum limit accuracy 
$\overline{(\delta p_z)^2}~\overline{(\delta z)^2}=1/4$.
(In our gedanken experiment, this corresponds to an uncertainty 
of $300~fm$ for position
and $300~nm/s$ for velocity.) Based 
on the master equation, we can predict the average position of the 
cantilever tip 
for its given initial state, depending on the spin state. If
the double uncertainty of the position is smaller than the separation 
between two
possible average positions, the cantilever tip will measure the state of 
the spin.\\ \ \\
After Fourier transformation, the ``cantilever part" of the density matrix 
is represented
by

\begin{equation}  
\hat\rho_{s,s'}(k,r,0)\propto\exp{\left[\displaystyle
 i p_0 r+ikz_0-r^2/4-k^2/4\right]}.
\label{rof}
\end{equation}

\section{\bf Solution for spin diagonal matrix elements} 

The equations for the characteristics of Eq. (\ref{eqrd}) are 
\begin{equation}
d\tau ={dr\over\beta r-k}={dk\over   r}=
{d\hat\rho_{s,s'}\over\biggl(-D\beta r^2+2 i  \eta
s r\biggr)  \hat\rho_{ss}}, 
\label{eqcd}
\end{equation}
or, explicitly 
\begin{equation} 
\begin{array} {lll}
   {\ds dr\over\ds d\tau} &=\beta r -k, \\ \ \\ 
   {\ds dk\over\ds d\tau}&=  r,\\ \ \\
  {\ds d\hat\rho_{s,s'}\over\ds d\tau}&=\biggl(-D\beta r^2+ 2 i  \eta
     s  r \biggr)~\hat\rho_{s,s'}.
\end{array}
\label{eqc1}
\end{equation}
From the first two equations in (\ref{eqc1}), one obtains
$$
{d^2k\over d\tau^2}-\beta {dk\over d\tau}+  k=0,
$$
which has the following general solution
\begin{equation} 
 k=e^{\beta \tau/2}\biggl(c_1\cos{\theta\tau}+c_2\sin{\theta\tau}\biggr),
\label{eqk}
\end{equation}
where
$
\theta=\sqrt{1 -{\beta^2\over 4}}
$. (Here we are considering the case $\beta < 2$ so  $\theta$
is a real number. The case $\beta > 2$ can also be solved analytically.)
Using the second equation in (\ref{eqc1})  one obtains:
\begin{equation}
r=e^{\beta \tau/2}\left[\left({\beta\over 2}\cos{\theta\tau}-\theta\sin{\theta
\tau} \right)c_1+ 
\left({\beta\over 2}\sin{\theta\tau}+\theta\cos{\theta\tau}\right)c_2\right].
\label{eqrr}
\end{equation}
Inverting Eqs. (\ref{eqk}) and (\ref{eqrr}) as a function of $c_1$ and
$c_2$ one obtain  the characteristic curves:
\begin{equation}
c_1=e^{-\beta\tau/2}( q_1k+ q_2r),
\label{ch1}
\end{equation}
 and
\begin{equation}
c_2=e^{-\beta\tau/2}( p_1k+ p_2r), 
\label{ch2}
\end{equation}
where the time dependent 
constants $q_1$, $q_2$, $p_1$ and $p_2$ have been defined as
\begin{equation}
\begin{array}{lll}
q_1&={\ds 1\over\ds\theta}\biggl({\ds\beta\over\ds 2}\sin{\theta\tau}+
\theta\cos{\theta
\tau}\biggr),\\ \\
q_2&=-{\ds 1\over\ds\theta} \sin \theta\tau,\\ \\
 p_1&={\ds 1\over\ds \theta}\biggl(-{\ds\beta\over\ds 2}\cos{\theta\tau}+
\theta\sin{\theta
\tau}\biggr),\\ \\
p_2&={\ds 1\over\ds\theta} \cos \theta\tau.\\
\label{qps}
\end{array}
\end{equation}
Substituting (\ref{eqrr}) into the third equation of (\ref{eqc1})
and integrating in time, one obtains:

\begin{equation} 
\hat\rho_{s,s}(k,r,\tau) \propto Q(c_1,c_2) \exp{\left[ i2\eta 
s(c_1g_1+c_2g_2)-D\beta
(c_1^2f_1+2c_1c_2f_3+c_2^2f_2)\right]},
\label{ss1}
\end{equation}
where the functions $f_i's$ and $ g_i's$  are defined as
\begin{equation}
\begin{array}{lll}
\displaystyle f_{_1}(\tau) &={e^{\beta\tau}\over 8}\biggl[\left(\beta+
{4\theta^2\over\beta}\right)+
\beta\cos{2\theta\tau}-2\theta\sin{2\theta\tau}\biggr],\\ \\
\displaystyle f_{_2}(\tau) &={e^{\beta\tau}\over 8}\biggl[\left(\beta+
{4\theta^2\over\beta}\right)-
\beta\cos{2\theta\tau}+2\theta\sin{2\theta\tau}\biggr],\\ \\
\displaystyle f_{_3}(\tau) &={e^{\beta\tau}\over 8}\biggl[2\theta
\cos{2\theta\tau}+\beta\sin{2\theta\tau}
\biggr],\\ \\
g_{_1}(\tau) &=e^{\beta\tau/2}\cos{\theta\tau},\\ \\
g_{_2}(\tau) &=e^{\beta\tau/2}\sin{\theta\tau}.
\end{array}
\label{feg}
\end{equation}
The arbitrary function $A$ which depends on the characteristics is
 determined by the initial density matrix 
$\hat\rho_{s,s}(k(0),r(0),0)$,
\begin{equation}
\begin{array}{lll}
A(c_1,c_2)&=\hat\rho_{s,s}\biggl(c_1,\frac{1}{2}\beta c_1+\theta c_2,0\biggr)
~\exp{\left[\displaystyle -2 i \eta s(c_1g_{10}+c_2g_{20})\right]}\\
&\quad\times\exp{\left[\displaystyle
D\beta(c_1^2f_{10}+2c_1c_2f_{30}+c_2^2f_{20})\right]},
\end{array}
\label{qqe}
\end{equation}
where $f_{i0}=f_i(0)$ and $g_{i0}=g_i(0)$.
From the initial density matrix (Eq. (\ref{rof})), we obtain  

\begin{equation}
\begin{array}{lll}
\hat\rho_{s,s'}(k,r,0)& \propto
\exp{\left\{ i\left[ \left( \frac{1}{2} p_0\beta +z_0 \right)c_1+
p_0\theta c_2\right]\right\}}\\ \\
&\times\exp{\left\{-\left[
\left({\beta^2\over 16}+{1\over 4}\right)c_1^2+
{\beta\theta \over 4} c_1 c_2 +
{\theta^2\over 4}c_2^2\right]\right\}}.
\end{array}
\label{rof1}
\end{equation} 
Substituting (\ref{qqe}) and (\ref{rof1}) into (\ref{ss1}) one obtains:

\begin{equation}
\begin{array}{lll}
    \hat\rho_{s,s}(k,r,\tau)&\propto\exp{\left\{\displaystyle i\left[
    \left( \frac{1}{2} p_0\beta +z_0+2\eta sG_1\right)c_1+
    \left(p_0\theta+2\eta s G_2\right)c_2\right]\right\}}\\ \\
&\quad\times\exp{\left\{\displaystyle-\left[
    \left({\beta^2\over 16}+{1\over 4}\right)c_1^2+
    {\beta\theta\over 4} c_1c_2+{\theta^2\over 4}c_2^2\right]\right\}}\\ \\
&\quad\times\exp{\left\{-D\beta( F_1 c_1^2+2 c_1 c_2 F_3+F_2 c_2^2)\right\}},
\end{array}
\label{e37}
\end{equation} 
where $F_i$ and $G_i$ are defined as
$$F_i(\tau)=f_i(\tau)-f_{i0},\quad\quad\quad\quad G_i(\tau)=g_i(\tau)-g_{i0}.
$$
Substituting in (\ref{e37}) the values of characteristics as a function of 
$k$ and $r$
(Eqs.(\ref{ch1}) and (\ref{ch2})), one obtains:

\begin{equation}
\hat\rho_{ss}(k,r,\tau) \propto\exp{\left[\displaystyle 
 -r^2 C_1 +irC_2 +
\left( iB_2 -r B_1 \right) k - \sigma_*^2 k^2 \right]},
\label{e38}
\end{equation}
where
\begin{equation}
\sigma_{*}^2 =e^{-\beta t} \left[ \left(
 {\beta^2\over 16}+{1\over 4}\right) q_1^2
+{\beta\theta \over 4} q_1 p_1 
+{\theta^2\over 4} p_1^2 
\quad+D\beta( F_1q_1^2+2q_1p_1F_3+F_2p_1^2) \right],
\label{esig}
\end{equation}

\begin{equation}
\begin{array}{lll}
 B_1&=e^{-\beta t}\biggl\{
\left({\beta^2\over 16}+{1\over 4}\right)2q_1q_2+
{\beta\theta\over 4}(q_1p_2+q_2p_1) +
{\theta^2\over 4}~2p_1p_2\\
&\quad+2D\beta [ 
F_1q_1q_2+(q_1p_2+q_2p_1)F_3+F_2p_1p_2]\biggr\},
\label{eb1}
\end{array}
\end{equation}

\begin{equation}
 B_2(s)=e^{-\beta t/2}\biggl[
\left(\frac{1}{2}p_0 \beta+z_0+2\eta s G_1\right)q_1+
\left(p_0 \theta+2\eta s G_2\right)p_1\biggr],
\label{eb2}
\end{equation}  

\begin{equation}
\begin{array}{lll}
C_1&= e^{-\beta t}\biggl[
\left({\beta^2\over 16}+{1\over 4}\right)q_2^2+
{\beta\theta \over\ 4 } q_2 p_2 +
{\theta^2\over4}p_2^2\\
&\quad +D\beta (F_1q_2^2+2q_2p_2F_3+F_2p_2^2)\biggr],\\
\label{ec1}
\end{array}
\end{equation}

\begin{equation}
  C_2(s)= e^{-\beta t/2}\biggl[
    \left(\frac{1}{2}p_0 \beta+z_0+2\eta s G_1\right)q_2+
    \left(p_0 \theta+2\eta s G_2\right)p_2\biggr].
\label{ec2}
\end{equation}
In Eqs. (31) and (33) the explicit dependence on $s$ is presented.
Performing  the inverse Fourier transform
one obtains 
\begin{equation}
\begin{array}{lll}
\rho_{1/2,1/2}(R,r,\tau)&={\ds |a|^2 \over\ds 
\sqrt{\pi}~\sigma_{*}}~
\exp{\left[\displaystyle - r^2 C_1+ ir C_2(1/2)\right]}\\
&\quad\quad\quad\quad\times\exp{\left[ ( -rB_1+iB_2(1/2)-iR)^2
/4\sigma^2_{*}\right]},\\ \ \\

\rho_{-1/2,-1/2}(R,r,\tau)&={\ds |b|^2 \over\ds 
\sqrt{\pi}~\sigma_{*}}~
\exp{\left[\displaystyle- r^2 C_1+ ir C_2(-1/2)\right]}\\
&\quad\quad\quad\quad\times\exp{\left[ ( -rB_1+iB_2(-1/2)-iR)^2/4
\sigma^2_{*}\right]}.\\
\label{efd}
\end{array}
\end{equation}
Eqs. (\ref{efd}) represent two squeezed Gaussians with modulus 

\begin{equation}
\begin{array}{lll}
|\rho_{1/2,1/2}(R,r,\tau)|&={\ds |a|^2 \over\ds 
\sqrt{\pi}~\sigma_{*}}~
\exp{\left[\displaystyle- r^2 (C_1- B_1^2/4\sigma^2_{*})\right]}\\
&\quad\quad\quad\quad\times\exp{\left[ -(B_2(1/2)-R)^2
/4\sigma^2_{*}\right]},\\ \ \\

|\rho_{-1/2,-1/2}(R,r,\tau)|&={\ds |b|^2 \over\ds 
\sqrt{\pi}~\sigma_{*}}~
\exp{\left[\displaystyle- r^2 (C_1- B_1^2/4\sigma^2_{*})\right]}\\
&\quad\quad\quad\quad\times\exp{\left[ -(B_2(-1/2)-R)^2
/4\sigma^2_{*}\right]}.
\label{efds}
\end{array}
\end{equation}
Fig. 2 shows schematically two peaks (seen from the top as ellipses) 
corresponding to
the two matrix elements $|\rho_{1/2,1/2}|$ and $|\rho_{-1/2,-1/2}|$. We denote
the centers of the ellipses, which lie on the diagonal $z=z'$, by $M_{++}$ 
and $M_{--}$, 
the semi-major axis by $\sigma_d$, the semi-minor axis by $\sigma_d'$ and the
distance between the centers by $\Delta_d$. The 
$\rho^{max}_{1/2,1/2}$ is located at $M_{++}=(r=0, R=B_2(1/2))$ or 
$z=z'=B_2(1/2)$, while the $\rho^{max}_{-1/2,-1/2}$
is at $M_{--}=(r=0, R=B_2(-1/2))$ or $(z=z'=B_2(-1/2))$.  The distance 
$ \Delta_d$
 is given by
\begin{equation}
\Delta_d = B_2(1/2)-B_2(-1/2).
\label{zmax}
\end{equation}
From Eqs. (\ref{efds}), we obtain $\sigma_d=\sqrt{2}~\sigma_*$, and

\begin{equation}
2{{\sigma'}_d}^2  = \displaystyle \frac{4\sigma_*^2}{4\sigma_*^2 C_1 - B_1^2}.
\label{scr}
\end{equation}
For a single spin measurement, the two peaks corresponding to 
$\rho^{max}_{1/2,1/2}$ and 
$\rho^{max}_{-1/2,-1/2}$ must be well separated. It follows that the condition 
$\Delta_d>2\sigma_d$ must be satisfied.\\ \\
First, we consider the case $\beta\tau\gg 1$ or $\tau\gg Q/\omega_c$, where 
$Q/\omega_c$ is
the time constant for the cantilever. In this case, we obtain two equilibrium 
positions for
the cantilever, when the transient process is over. We have $\Delta_d=2\eta$ 
and $\sigma_d=\sqrt{D}$.
The value $\sigma_d=\sqrt{D}$ is the thermodynamical uncertainty in the 
cantilever position caused
by the thermal noise. 
The two equilibrium positions can be distinguished if $\eta>\sqrt{D}$ or 
$T<F^2/k_{_B}k_c$, where
$F=\mu_{_B}|\partial B_z/\partial z|$ is the magneto-static force between 
the ferromagnetic particle
and the paramagnetic atom. The last expression exactly coincides with 
formula (1).\\ \ \\
Next, we consider the initial transient process after the instant ($t=0$) 
at which the paramagnetic 
spin has been transferred into the superpositional state. 
For $\beta\tau\ll 1$, we have

\begin{equation}
\Delta_d=4\eta\sin^2\frac{\tau}{2},\quad \sigma_d=\left[1/2+D\beta\tau-
D\beta\cos\tau\sin^3\tau\right]^{1/2}.
\end{equation}
This expression for $\Delta_d$ describes the oscillating distance between 
the two peaks. It corresponds to
initial vibrations of two classical oscillators near their equilibrium 
positions $z=\eta$ and 
$z=-\eta$.
The distance between them is
given by $\Delta_d$. (For our gedanken experiment the maximum value of 
$\Delta_d$ is $0.24~nm$.)
The formula for $\sigma_d$ contains three terms. The first term, $1/2$, 
corresponds
to the quantum dispersion of the initial wave function. The second term, 
$D\beta\tau$, describes the
initial diffusion of an ensemble of oscillators. Formally, setting 
$\tau\sim 1/\beta$, we can estimate
the final dispersion $\sigma_d=\sqrt{D}$, which corresponds to 
thermodynamical vibrations of the
cantilever tip. The third term describes insignificant oscillations 
with small amplitude, $D\beta$.\\ \\ 
Note that the condition for distinguishing two cantilever positions at 
the beginning of the 
transient process
is much less restrictive than the corresponding condition for the 
equilibrium positions
 at $\beta\tau>>1$.
Indeed, after the first half-period ($\tau=\pi$), we have $\Delta_d=4\eta$ 
and 
$\sigma_d=(1/2+ \pi D\beta)^{1/2}$. Taking into account that $\beta\ll 1$, 
the condition for 
distinguishing two positions, 
$\eta>(1/2+\pi D\beta)^{1/2}$, is much easier than $\eta>\sqrt{D}$. In our 
gedanken 
experiment the condition for distinguishing the two positions for the 
transient process is
\begin{equation}
T<T_{max}=\frac{4}{\pi}\frac{QF^2}{k_{_B}k_c}=14~K,
\label{eee}
\end{equation}
compared with $T<T_{max}=1.7~mK$ for the static Stern-Gerlach effect. 
This estimate seems to be too 
optimistic. It is connected with the very small distance ($5~nm$) between 
the ferromagnetic particle
and the paramagnetic atom. If we increase this distance to $50~nm$, the 
temperature $T_{max}$ drops
from $14~K$ to $1.1~mK$. Note that expression (\ref{eee}) coincides exactly 
with our preliminary
estimates (2).\\ \ \\
\noindent
The condition $\Delta_d>2\sigma_d$ is satisfied for the first time at

\begin{equation}
\tau=\tau_0\approx 2^{1/4}/\sqrt{\eta}.
\end{equation}
This expression is valid if $\eta>>1$ and $\eta>>(D\beta)^2/\sqrt{8}$. 
For our gedanken experiment
we have $\eta=144$, $D=1.25\times 10^7T$ ($T$ is the temperature in Kelvin), 
$\beta=1.5\times 10^{-4}$, and $\tau_0=0.1$. Thus, the above conditions 
are both satisfied. The value of 
$t_0=\omega_c\tau_0$ is approximately $9.3~\mu s$.

\section{\bf Solution for off-diagonal spin matrix elements}
The equations for the characteristics   are now given by
\begin{equation}
d\tau={dr\over \beta r-k}={dk\over  r-4\eta s}={d\hat\rho_{s,-s}\over
-D\beta r^2 \ \hat\rho_{s, -s}},
\label{chnd}
\end{equation}
or

\begin{equation}
\begin{array}{lll}
    {\ds dr\over\ds d\tau}&=\beta r-k,\\ \ \\
    {\ds dk\over\ds d\tau}&=  r-4\eta s,\\ \ \\
    {\ds d\hat\rho_{s, -s}\over\ds d\tau}&=-D\beta r^2~\hat\rho_{s,-s}.\\
\end{array}
\label{chnd2}
\end{equation}
The solutions of the first two equations of (\ref{chnd2}) are 

\begin{equation}
\begin{array}{lll}
k&=e^{\beta \tau/2}\biggl(c_1\cos{\theta\tau}+c_2\sin{\theta\tau}\biggr)+
4 \beta
\eta s \\
r&=e^{\beta\tau/2}\biggl[\left({\beta\over 2}\cos{\theta\tau}-\theta
\sin{\theta
 \tau}\right)c_1+ \left({\beta\over 2}\sin{\theta\tau}+
\theta\cos{\theta\tau}\right)c_2\biggr]+ 4 \eta s.\\
\end{array}
\label{solnd2}
\end{equation}
Following the same steps as above we obtain for the Fourier transform:

\begin{equation}
\begin{array}{lll}
\hat\rho_{1/2, -1/2}(k,r,\tau)& \propto A(c_1,c_2)  
\exp{\left\{\displaystyle -D\beta\biggl[
    f_1c_1^2+2c_1c_2f_3+f_2c_2^2\biggr]\right\}}\\ 
&\quad\quad\quad\times\exp{\left\{ -D\beta\biggl[4 \eta  
(g_1c_1+g_2c_2)+ 4\eta^2\tau\biggr]\right\}},
\end{array}
\label{rndf}
\end{equation}
where we fix $s=1/2$ (changing sign of $s$ 
corresponds to a
change of sign of $\eta$, see Eq.(\ref{chnd2}), therefore
the case $s=-1/2$ can be easily
obtained).
The functions $f_i$ and $g_i$ have been defined as above,
and $c_1$ and $c_2$ are new characteristic curves given by
\begin{equation}
\begin{array}{lll}
c_1 &=e^{-\beta\tau/2}( q_1 k+ q_2 r+ \eta q_3 ),\\
c_2 &=e^{-\beta\tau/2}( p_1 k+ p_2 r+ \eta p_3).\\
\label{ccnd}
\end{array}
\end{equation}
Here, $q_1, q_2, p_1, p_2$ are defined by Eqs. (\ref{qps}) and $q_3,
p_3$ are given by

\begin{equation}
\begin{array}{lll}
q_3 &={\ds {2} \over\ds \theta}\biggl[
-\beta \biggl({\beta\over 2}\sin{\theta\tau}+
\theta\cos{\theta\tau}\biggr)+\sin{\theta\tau} \biggr],\\
p_3 &={\ds {2}\over\ds \theta}\biggl[
\beta \biggl({\beta\over 2}\cos{\theta\tau}-
\theta\sin{\theta\tau}\biggr)-\cos{\theta\tau} \biggr].\\
\label{qp0}
\end{array}
\end{equation}
With the same initial condition Eq. (\ref{rof}), we can determine the 
function  $A (c_1, c_2 )$ and obtain

\begin{equation}
\begin{array}{lll}
&\hat\rho_{1/2,-1/2}(k,r,\tau)\propto\hat\rho_{1/2,-1/2}\biggl(c_1+2 
\beta \eta,
{1\over 2 }\beta c_1
+ \theta c_2+2\eta ,0 \biggr)\\
 &\quad \times\exp{\left\{\displaystyle -D\beta\biggl[
F_1c_1^2+2c_1c_2F_3+F_2c_2^2+4\eta (G_1c_1+G_2c_2)+4 \eta^2\tau\biggr]
\right\}},
\label{afu}
\end{array}
\end{equation}
where $F_i(\tau)$ and $G_i(\tau)$ are defined as above.
By substituting the initial condition (\ref{rof}) we have

\begin{equation}
\begin{array}{lll}
&\hat\rho_{1/2,-1/2}(k,r,\tau)\propto\exp{\left\{ \displaystyle i\biggl[
(\frac{1}{2}p_0 \beta+z_0 ) c_1+p_0\theta c_2 +
2\eta(p_0+z_0\beta) \biggr]\right\}}\\
&\quad\quad\quad\times \exp{\left\{\displaystyle -\biggl[ 
\left( {\beta^2\over 16}+
{1\over 4} \right) c_1^2+ {\beta\theta \over 4} c_1 c_2+
{\theta^2 c_2^2 \over 4}  \biggr]\right\}}\\
&\quad\quad\quad\times \exp{\left\{\displaystyle -\biggl\{4\eta
\left[\left({3\beta\over 8}\right)
c_1+{\theta\over 4}c_2\right]+ \eta^2\left(
{1}+\beta^2\right)\biggr\}\right\}}\\
&\quad\quad\quad\times \exp{\left\{\displaystyle -D\beta \biggl[ 
F_1c_1^2+2c_1c_2F_3+F_2c_2^2+
4\eta s (G_1c_1+G_2c_2)+\eta^2 \tau\biggr]\right\}},\\
\label{rof22}
\end{array}
\end{equation}
which can be written as 

\begin{equation}
\begin{array}{lll}
\hat\rho_{1/2,-1/2}(k,r,\tau)& \propto\exp{\left[\displaystyle 
 -r^2 C_{12}-r\eta C_{11}- \eta^2 C_{10} +i r C_{21}+ i\eta C_{20}\right]}\\
&\quad\times\exp{\left[\left(i B_{20} -r B_{11} - \eta B_{10} \right) k - 
\sigma_*^2 k^2\right] },
\end{array}
\label{e38nd}
\end{equation}
where $\sigma_*$ is given by (\ref{esig}) and
 
\begin{equation}
\begin{array}{lll}
C_{12}  &= e^{-\beta\tau} \biggl[ \left({\ds\beta^2\over\ds 16}+
{\ds 1\over\ds  4} +D\beta F_1\right) q_2^2 +\left({\ds\beta\theta\over\ds 4}
+2D\beta F_3\right)q_2 p_2 + \left({\ds\theta^2\over\ds 4}+D\beta F_2
\right)p_2^2 \biggr],\\ \ \\
C_{11}  &= e^{-\beta\tau} \biggl[ \left({\ds\beta^2\over\ds 16}+
{\ds 1\over\ds  4} +D\beta F_1\right)2 q_2 q_3 +
\left({\ds\beta\theta\over\ds 4} +2D\beta F_3\right)(q_2 p_3 + p_2 q_3)
\biggr]\\
&\quad + e^{-\beta\tau} \biggl[ 
\left({\ds\theta^2\over\ds 4 }+D\beta F_2 \right)2 p_2 p_3 \biggr] \\
&\quad+4 e^{-\beta\tau/2} \biggl[ \left({\ds 3\beta\over\ds 8 } +  
D\beta G_1 \right)q_2 +
 \left({\ds\theta\over\ds 4}   
 + D\beta G_2 \right)p_2 \biggr],\\ \ \\
C_{10}  &= e^{-\beta\tau} \biggl[ \left({\ds\beta^2\over 16}+
{\ds 1\over\ds  4} +D\beta F_1\right) q_3^2 +\left({\ds\beta\theta\over\ds 4}
+2D\beta F_3\right)q_3 p_3 + \left({\ds\theta^2\over\ds 4 }+D\beta F_2
\right)p_3^2 \biggr]    \\
&\quad+4 e^{-\beta\tau/2} \biggl[ \left({\ds 3\beta\over\ds 8} +  
D\beta G_1 \right)q_3 +
 \left({\ds\theta\over\ds 4 }   
 + D\beta G_2 \right)p_3 \biggr]\\
&\quad+ 4   \left({\ds 1\over\ds 4}+ \frac{\ds\beta^2}{\ds 4}
+D\beta\tau \right),\\ \ \\
C_{21}  &= e^{-\beta\tau/2}\biggl[ \left( \frac{\ds p_0\beta}{\ds 2}+ z_0
\right)q_2
+ p_0\theta p_2 \biggr],   \\ \ \\
C_{20}  &= e^{-\beta\tau/2}\biggl[ \left( \frac{\ds p_0\beta}{\ds 2}+ z_0
\right)q_3
+ p_0\theta p_3 \biggr]+ 2 \left( p_0 + z_0 \beta \right),     \\ \ \\
B_{11}  &= e^{-\beta\tau} \biggl[ \left({\ds\beta^2\over 16}+
{\ds 1\over\ds  4} +D\beta F_1\right) 2 q_2 q_1 
+\left({\ds\beta\theta\over\ds 4} +2D\beta F_3\right)
(q_1 p_2 +q_2 p_1)\biggr]\\
&\quad+ e^{-\beta\tau} \biggl[ \left({\ds\theta^2\over\ds 4}+ D\beta F_2
\right) 2 p_2 p_1  \biggr],    \\ \ \\
B_{10}  &= e^{-\beta\tau} \biggl[ \left({\ds\beta^2\over\ds 16}+
{\ds 1\over\ds  4} +D\beta F_1\right) 2 q_3 q_1 
+\left({\ds\beta\theta\over\ds 4} +2D\beta F_3\right)
(q_1 p_3 +q_3 p_1)\biggr]\\
&\quad+ e^{-\beta\tau} \biggl[ \left({\ds\theta^2\over\ds 4}+D\beta F_2
\right)  2 p_3 p_1  \biggr]    \\  
&\quad+ 4  e^{-\beta\tau/2} \biggl[ \left({\ds 3\beta\over\ds 8 } +  
D\beta G_1 \right)q_1 +
 \left({\ds\theta\over\ds 4}   
 + D\beta G_2 \right)p_1 \biggr],\\ \ \\
B_{20}  &= e^{-\beta\tau/2} \biggl[ \left(\frac{\ds 1}{\ds 2}p_0 \beta + z_0 
\right)q_1 +p_0 \theta p_1 \biggr].
\label{cabnd}
\end{array}
\end{equation}

Performing the inverse Fourier transform we obtain

\begin{equation}
\begin{array}{lll}
\rho_{1/2,-1/2}(R,r,\tau) &= 
\displaystyle \frac{a b^*}{\sqrt{\pi}\sigma_*} 
\exp{\left[\displaystyle   -r^2 C_{12}-r\eta C_{11}-\eta^2 C_{10} +i r C_{21}+ 
i\eta C_{20}\right] } \\
&\quad\times\exp{\left[\displaystyle
\left( -r B_{11} - \eta B_{10} +i B_{20} -i R \right)^2/ 4 
\sigma_*^2\right] },\\
\rho_{-1/2,1/2}(R,r,\tau) &=
\displaystyle \frac{a^* b}{\sqrt{\pi}\sigma_*} 
\exp{\left[\displaystyle   -r^2 C_{12}+r\eta C_{11}-\eta^2 C_{10} +i r C_{21}-
i\eta C_{20}\right] } \\
&\quad\times\exp{\left[\displaystyle
\left( -r B_{11} + \eta B_{10} +i B_{20} -i R \right)^2/ 4 \sigma_*^2\right] }.
\label{aftnd}
\end{array}
\end{equation}
Taking the modulus one obtains 

\begin{equation}
\begin{array}{lll}
|\rho_{1/2,-1/2}(R,r,\tau)| &=
\displaystyle \frac{|a b^*|}{\sqrt{\pi}\sigma_*} \ e^{\xi \eta^2}
\ e^{\displaystyle   - (r + r_0\eta)^2/2\tilde\sigma^2} 
\ e^{\displaystyle -( B_{20} - R )^2/ 4 \sigma_*^2 },\\
|\rho_{-1/2,1/2}(R,r,\tau)| &=
\displaystyle \frac{|a^* b|}{\sqrt{\pi}\sigma_*} \ e^{\xi \eta^2}
\ e^{\displaystyle   - (r - r_0\eta)^2/2\tilde\sigma^2}
\ e^{\displaystyle -( B_{20} - R )^2/ 4 \sigma_*^2 },
\label{aftndm}
\end{array}
\end{equation}
where

\begin{equation}
\begin{array}{lll}
\tilde\sigma^2 &=  \frac{\ds 2\sigma_*^2}{\ds 4\sigma_*^2 C_{12} -
B_{11}^2},\\ \\
r_0&=  \frac{\ds 2\sigma_*^2 C_{11}-B_{11}B_{10}}
{\ds 4\sigma_*^2 C_{12} - B_{11}^2},\\ \\
\xi&=  \frac{\ds B_{10}^2}{\ds 4\sigma_*^2}- C_{10}-\frac{\ds r_0^2}
{\ds 2\tilde\sigma^2}.
\label{wndm}
\end{array}
\end{equation}
The maxima are located at 
$(R=B_{20}, r=-\eta r_0)$ for $|\rho_{1/2,-1/2}|$)
and at $(R=B_{20}, r=\eta r_0)$ for  $|\rho_{-1/2,1/2}|$.
In $(z,z')$ coordinates this corresponds to 
$M_{+-}= (z=B_{20} -\eta r_0/2, z'=B_{20}+\eta r_0/2)$
for $|\rho_{1/2,-1/2}|$ and 
$M_{-+}= (z=B_{20} +\eta r_0/2, z'=B_{20}-\eta r_0/2)$
for $|\rho_{-1/2,+1/2}|$, so that the distance between them
is given by $\Delta_{nd} = \sqrt{2} \eta |r_0|.$
Next, we consider the quadratic form 
$ (r\pm r_0\eta)/2\tilde\sigma^2 + (B_{20}-R)^2/4\sigma_*^2$
in the $(z,z')$ plane. Straightforward calculations show that this is an 
ellipse
whose semi-axes are respectively given by $\tilde\sigma$ (across the diagonal)
and $2\sqrt{2}\sigma_*$ (along the diagonal).
The centers of the peaks $M_{+-}, M_{-+}$ are symmetric with respect to
the diagonal line $z=z'$.\\ \\
\noindent
The most remarkable difference compared with the diagonal case, is 
the presence of irreversible decoherence. Indeed, the heights of the
peaks are exponentially reduced in time  by the damping factor 
$ \sim 
\exp(-4\eta^2 D \beta\tau)$. This, in turn, defines a characteristic
time scale of decoherence: $\tau_d=1/4\eta^2D\beta$. This formula exactly 
coincides with
the expression derived in \cite{g}, based on a semi-qualitative analysis. At 
time $\tau=\tau_0$,
when two diagonal peaks are clearly separated, the damping factor is 
$4\eta^2D\beta\tau_0$. 
We expect to observe the coherence between the two peaks (MSCS) if this 
factor is not
much more than one unit. Thus, using the expression for $\tau_0$ from Eq. 
(40), we can estimate 
the condition for the quantum coherence as 
$D\beta\eta^{3/2}<1$, or

\begin{equation}
T<T_{max}={\ds Q\over\ds k_{_B}}{\ds (\hbar\omega_c)^{7/4}k_c^{3/4}\over\ds 
F^{3/2}}.
\end{equation}
For our gedanken experiment the value of $T_{max}$  is approximately  
$3\times 10^{-7}~K$.\\ \ \\
Now, we will check the validity of our estimate for the parameters chosen 
for our gedanken experiment. 
Our solution is valid if 
$\eta>>(D\beta)^2/\sqrt{8}$. Setting $T=T_{max}$ or $D\beta\eta^{3/2}=1$, 
we obtain 
$\eta^4>>1/\sqrt{8}$, which is definitely true, assuming $\eta>>1$. Next, 
the condition of
the validity of the high temperature approximation is $D>>1$. For $T=T_{max}$, 
it follows that
$\eta^{3/2}<<Q$. This inequality is roughly satisfied ($\eta^{3/2}=1700$, 
$Q=6700$). Finally,
as we mentioned in the Introduction, the master equation fails at times 
$t\leq\hbar/k_{_B}T$. Thus,
the time considered, $\tau_0=2^{1/4}/\sqrt{\eta}$, must be much greater 
than $1/D$, which is 
definitely wrong. Thus, our condition (54) for the creation of MSCS is 
not justified
for the parameters considered. \\ \ \\ 
Next, we discuss at what values of the parameters a MSCS can be generated. 
First, we emphasize the qualitative difference between  two conditions. 1) 
The condition
for distinguishing two positions of the cantilever. 2) The condition for 
distinguishing two position
of the cantilever and the coherence between these two positions.
The first condition is relatively simple: $T<T_{max}=F^2/k_{_B}k_c$ for 
equilibrium positions 
($\tau>>Q$) and 
 $T<T_{max}=4QF^2/\pi k_{_B}k_c$ for $\tau=\pi$. The obvious way to 
increase $T_{max}$ is by decreasing
the spring constant $k_c$ or increasing the magneto-static force $F$. 
For $\tau=\pi$, one additional 
way  is to increase the quality factor $Q$. 
Condition 2) for generating an MSCS can be satisfied at temperature 
$T<T_{max}=\hbar^{7/4}Qk_c^{13/8}/k_{_B}F^{3/2}m^{7/8}$,
where $m$ is the effective mass  of the cantilever, $m=k_c/\omega_c^2$.
At $T=T_{max}$, the MSCS will be generated if we satisfy the inequalities 
$1\ll\eta^2\ll Q$, or

\begin{equation}
1<<{\ds m F^2\over\ds \hbar k_c^{3/2}}\ll Q.
\label{eqq}
\end{equation}
One can see that the regime considered in
our paper does not allow free manipulation of any parameter but $Q$. 
Increasing $Q$,  we can
satisfy  the second inequality and, at the same time, increase $T_{max}$. 
In our gedanken experiment,
the tenfold increase of $Q$ ($Q=67000$) provides the validity of the 
right-hand inequality in
 (\ref{eqq}) and increases $T_{max}$ to $3~\mu K$.

\section{\bf Conclusion}

We have shown that the maximum temperature for a single spin measurement 
in MFM can be increased
by a factor of $Q$  if one utilizes the initial transient process instead 
of the static
displacement of the cantilever tip. We have obtained an exact analytical 
solution 
of the master equation, which describes the $Q$-times magnification of 
the maximum temperature.
In addition, we have found the conditions for generation of macroscopic 
Schr\"odinger cat state in MFM.
\\ \ \\ 
\section{Acknowledgments}

The work of GPB was supported by the Department of Energy (DOE)
under the contract W-7405-ENG-36. GPB and VIT  thank 
the National Security Agency (NSA), the Advanced Research and 
Development Activity (ARDA), and DARPA (MOSAIC) for partial support.

\vfil\eject

\vfil\eject

%%%%%%%%%%%%%%%%%%%%%%%%%%%%%%%%%%%%%%%%%%%%%%%%%%%%%%%%%%%%%%%%%

\begin{figure}
\begin{center}
\epsfxsize 14cm \epsfbox{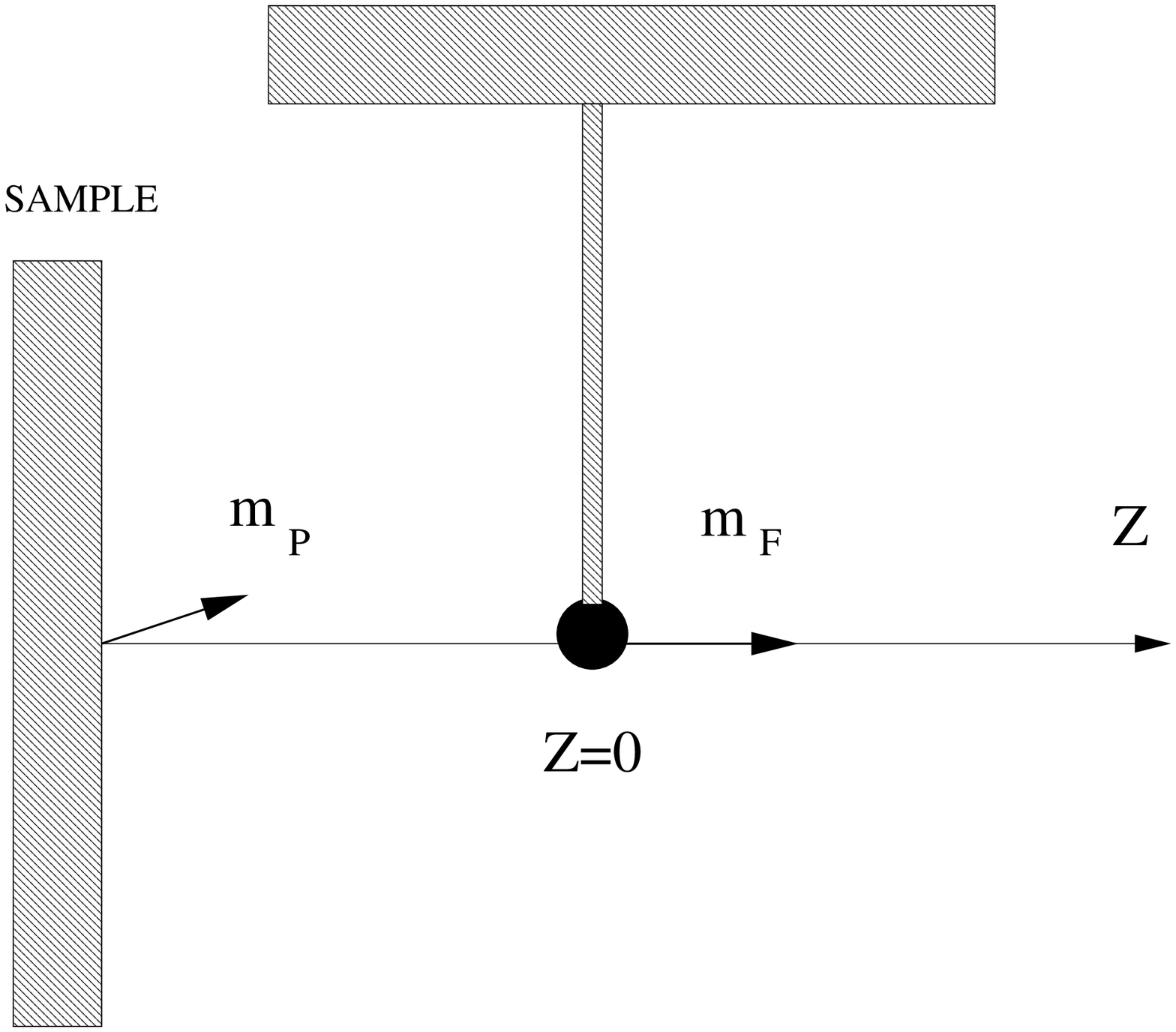}
\caption{
Geometry of the proposed gedanken experiment. ${\bf m}_{_F}$ and ${\bf m}_p$ 
are 
the magnetic moments of the ferromagnetic particle and paramagnetic atom.
}
\label{1}
\end{center}
\end{figure}

\begin{figure}
\begin{center}
\epsfxsize 14cm \epsfbox{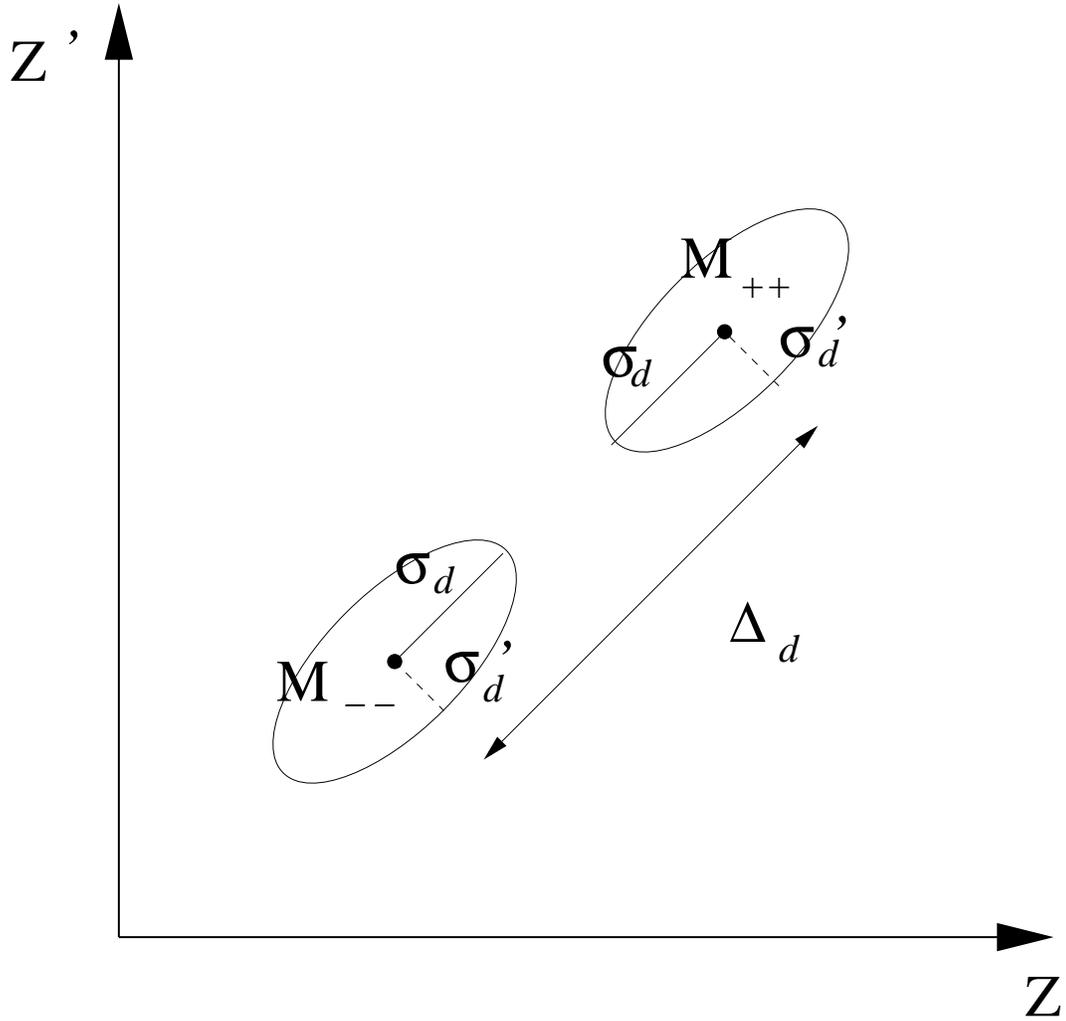}
\caption{ 
Schematic view of the Gaussians representing the diagonal elements 
$|\rho_{-1/2,-1/2}|$
and $|\rho_{1/2,1/2}|$ (seen from the top) in the ($z,z'$) plane. We show 
the centers
$M_{--}$ and $M_{++}$, the variances $\sigma_d'$ (transverse) and $\sigma_d$ 
(parallel), and
the distance between diagonal centers $\Delta_d$.
}
\label{xxx2}
\end{center}
\end{figure}

\end{document}